\documentclass[12pt,preprint]{aastex}

\usepackage{color}

\def\BB{{\bf {B}}}

\def\xx{{\bf {x}}}

\begin{document}

\title{Current singularities in line-tied 3D magnetic fields}

\author{I.~J.~D.~Craig}
\affil{Department of Mathematics, University of Waikato, New Zealand}
\email{i.craig@waikato.ac.nz}

\author{D.~I.~Pontin}
\affil{Division of Mathematics, University of Dundee, UK}
\email{dpontin@maths.dundee.ac.uk}

\begin{abstract}

This paper considers the  current distributions that derive
from finite amplitude perturbations of  line-tied magnetic
fields comprising hyperbolic field structures.  The initial
equilibrium on which we principally focus is a  planar magnetic $X$-point threaded by 
a uniform axial field.    This field is line-tied on all surfaces but subject to
3D disturbances that  alter the  initial  topology.   Results of ideal relaxation simulations are
presented  that  illustrate how intense current structures form that can be
related,  through  the influence of line-tying,    to the 
quasi-separatrix  layers (QSL's)  of the initial configuration.
It is demonstrated that the location within the 
QSL that attracts the current, and its scaling properties, are
strongly dependent on the relative dimensions of the QSL with respect to the line-tied boundaries.
These results are contrasted with the behaviour of a line-tied 3D field
containing an isolated null point. In this case it is found that
  the a dominant current always forms at the null, but that the 
collapse is inhibited when the null is closer to a line-tied boundary.
 
\end{abstract} 

\section{Introduction}

In a series of  previous studies the development of ideal current singularities
in magnetic null points has been investigated using a mixture of analytic arguments
and numerical experiments \citep{craiglitvinenko2005,pontincraig2005}.    It is well known  that when
magnetic equilibria of complex topology are perturbed, currents will be generated that
can only be dissipated by magnetic reconnection \citep{priest2000}.   Yet,  in a highly
conducting magnetic plasma such as the solar corona,  reconnection
can  be effective only in highly localized regions of strong current
density.      In the case of 3D magnetic null points,  
currents can form locally at the null in response to a magnetic
collapse in the weak-field region,  often forming quasi one-dimensional
current sheets or quasi-cylindrical current tubes.  The geometry of
the current distributions derive from  ``spine" and ``fan" structures
that form the skeleton of the null---essentially the the eigenstructure of the 3D field---and these
different current structures define the forms the  
reconnection can take \citep[e.g.][]{craig1996,pontin2011b}.    
Accordingly,  a useful route for understanding reconnection is  to
examine the near-singular current distributions that derive from 
perturbing 3D magnetic fields. 

In a previous paper  \cite{pontincraig2005} used an ideal Lagrangian 
magneto-frictional method to examine the singular current
distributions that derive from perturbing 2D and 3D compressible magnetic 
null points.   One of the aims of that study was to quantify
the extent to which finite gas pressure could inhibit the development of the 
current singularity,  and thus presumably slow down the reconnection rate in the realistic case of a small but finite resistivity.   Our present 
purpose is to complement this  work  by examining the stabilising role 
of line-tying on the formation of current singularities.  To do this 
we initially consider perturbing a 2D planar $X$-point threaded by a uniform perpendicular 
field $ B_z $.    The  equilibrium field has an ignorable
$z$-coordinate,   and contains no neutral point,   but 
perturbations  which alter the magnetic topology naturally lead to singular 3D current sheets.    We consider
two questions:   first,  how effective is line-tying the axial field
in determining the form and strength of the current singularity;  second,   to what extent can these
results be applied to line-tied fields containing 3D nulls.               

In \S 2 we introduce the line-tied $X$-point geometry that forms
the basis of our initial investigation.    We point out that the
presence  of a line-tied,  axial field
component introduces quasi-separatrix layers (QSL's) that fundamentally alter 
the causal properties of the configuration.   Relaxation simulations in 3D show that
line-tying  manifests itself  by inhibiting the blow-up of the current
density around the axis of the QSL.   In \S 3 corresponding relaxation   
computations are performed  for the case of a  3D null-point field.    The relaxed
current distribution now highlights a competition that develops 
between currents localized at the null and those localized at the
line-tied footpoints. In this case
line-tying is  effective in slowing the divergence of the 
``reconnective currents''  that  localize around the null.   Our
conclusions are presented in \S 4.

\section{$X$-point threaded by a uniform axial field}

\subsection{The equilibrium configuration}  
          
The simplest magnetic configuration of interest is the 2D potential
$X$-point $ {\bf B} = \nabla \psi \times \hat {{\bf z}} $ defined by
the planar flux function $ \psi = B_0 \, x y $.   We assume the gas pressure
is negligible but impose line-tying on the boundary of  the region  
{$ -1\le x,y \le 1 $.}

Consider adding  a  perturbation $ {\bf b}  = b_0 x  \,  \hat {\bf y}$ to
this field so that 
\begin{equation} 
\psi(x,y)  = B_0  \, x  y  - b_0  \, \frac {x^2}{2}.  
\label{twod}  
\end{equation}
The separatrices---field lines threading the null that delineate
regions of distinct flux---are no longer at right-angles but are tilted through the additional angle 
\begin{equation} 
\tan \theta = \frac {b_0} {2 B_0}.   
\label{sep}  
\end{equation}
The equilibrium can be regained only by flux transfer across the separatrices.  In a weakly resistive plasma, this involves
an implosion of the disturbance towards the $X$-point, generating a current sheet at which magnetic reconnection occurs
{\citep{mcclymont1991,hassam1992}}.     
Note that if resistive effects are turned off,  but some other form of damping is present,  the
final configuration  comprises  a singular distribution of 
current  in the weak field regions close to  the null {\citep{green1965,syrovatskii1971}.             

Now suppose  that the initial equilibrium is threaded by a uniform 
axial field  $ B_z $ of infinite length:  
\begin{equation} 
 {\bf B}_E = \nabla \psi \times \hat {{\bf z}}  + B_z {\hat {\bf z} },
 \qquad \psi = B_0 \,  x y 
\label{equi}
\end{equation}  
The null point is removed---fundamentally altering  the causal properties of the
configuration since the Alfv\'en speed is now non-vanishing---but a disturbance typified by (\ref{twod})  still alters the topology.   
That is,   by projecting field lines onto a surface $ z =\, $constant,   effective ``separatrices'' can 
be defined that delineate regions of distinct flux.   Field lines on
the effective separtrices extend to infinity in one direction 
rather than connecting back to the boundary as in the case of the planar
$X$-point.   The presence  of the $ B_z $ field introduces  an
extra pressure (of magnitude $ {B_z}^2 /2$)  that  resists the subsequent implosion.   Detailed computations for small 
amplitude perturbations  confirm that the magnetic pressure weakens the 
current singularity but cannot stop it forming  
\citep{mcclymont1996,craiglitvinenko2005}.
This is not surprising given that the problem
maintains the symmetry $ \partial_z = 0 $   in the absence of
line-tying.   In this case  finite  $ B_z $   leads to an  irrotational force $  \nabla ({B_z}^2 /2) $,
acting much like gas pressure,   that cannot balance the Lorentz
force that  drives the implosion.   This argument fails only under
highly restrictive conditions---for instance, one-dimensional current
layers---which are unlikely to apply in  the  present study.        

The symmetry $ \partial_z = 0  $ is broken when   the axial field  is line-tied on upper and  lower surfaces, $z=\pm z_m$, say.   This
leads to an equilibrium that  has been studied for a number of years in the context of 
current sheet formation and three-dimensional reconnection.   More
specifically,   when contained within a bounded domain, the field (\ref{equi})
contains a new topological feature,   the so-called
quasi-separatrix layer (QSL) \citep{titov2002,titov2007}.   The QSL is centred
on the $z$-axis and extends along the $y$-axis on the 
lower boundary and along the $x$-axis on the upper boundary (assuming
$B_0,B_z>0$) \citep{priest1995}.   More generally QSL's can be related
to geometrical squashing factors associated with  field line mappings
in line-tied magnetic configurations \citep{titov2007}.    We note
that in this fully three-dimensional configuration there no 
longer exist regions of distinct magnetic flux---the boundaries
between which correspond to discontinuities in the field line
mapping---but the QSL delineates a thin layer across which this
mapping has a strong variation.

Of interest to the present study is the current build-up due to driving
motions on the field boundaries.
Although  QSL's  provide
natural surfaces for current accumulation in 3D line tied configurations
such as (\ref{equi}),   there is no single point within the QSL that
provides a focus for the current localization that derives from
footpoint driving.   \cite{inverarity1997} demonstrated linearised solutions in which the field evolved 
through a sequence of force-free states in response to a slow driving
flow on the $z$-boundaries.  For some driving flows they found a
current singularity of a logarithmic type.     Of more direct
relevance are  the resistive MHD simulations of  \cite{galsgaard2000}
who subjected the lower $z$-boundary to  
various driving motions -- including a simple shear flow that leads
locally to a separatrix displacement as proposed in equation
(\ref{twod}).  A  strong current enhancement within the QSL was
obtained.  

Below we point out that the current structures obtained by
\cite{galsgaard2000}  bear a strong resemblance to the ``relaxed''
magnetic structures obtained in the present simulations.    However,
the  form and intensity of the current structure obtained is likely to
be strongly dependent on the form of perturbation applied to the
system,  as observed by \cite{inverarity1997,galsgaard2000}.   In what
follows, we focus on what   we call ``topological disturbances", that
is perturbations that disturb the location of the 
QSL (or in Section \ref{3dsec} the true separatrices).

\subsection{Magneto-frictional relaxation}

Current structures are obtained by adding 
topological disturbances onto the  line-tied equilibrium field
 (\ref{equi})  and  following the subsequent evolution using  an
ideal Lagrangian  scheme \citep{craig1986}. 
The present  version of the code  uses fourth-order
spatial differencing within a rectangular region $ -1 \le x,y \le 1 $ of height $ 2 z_m $ centred on
the origin.     Fluid particles on the boundaries are held fixed but
those in the interior---driven by the $ {\bf J} \times {\bf B}$ forces and subject to frictional damping---are
followed until the computed Lorentz forces are negligible.   The code
features implicit time-stepping,  satisfies $ \nabla \cdot {\bf B} =
0$ to machine accuracy,    and is unconditionally stable.
Effects due to finite gas pressures are assumed negligible.    

The Lagrangian scheme works on the principle that fluid line elements
evolve in the same way as $ {\bf B}/ \rho $ (where $ \rho $ is the
local mass density).     Since  fluid particle displacements $ {\bf \xi}(x,y,z) $ are the
 primary variables,  we choose initial conditions according to
 the induction equation for small perturbations,  namely  
\begin{equation} 
 {\bf b}(x,y,z) =  \nabla \times  ({\bf \xi} \times  {\bf B}_E) 
\label{icond}
\end{equation}  
where $  {\bf B}_E $ is given by  (\ref{equi}).   A 
displacement  that reproduces  (\ref{twod})  is provided  by $ {\bf \xi}  =
b_0 x/ (2 B_0) \, {\hat {\bf y}} $.     This form is conveniently
extended  into the computational 3D domain by assuming separable
functions of $y$ and $z$ that vanish on the boundaries.
It should be stressed that,  although perturbations constructed in
this way are expected to provide a divergent current layer,  the
frictional relaxation could,   at least  potentially,   evolve towards
 a force-free equilibrium that involves non-singular  current  structures.
 
We begin by considering the result of applying 
a perturbation of the form  
\begin{equation} 
 {\bf \xi}(x,y,z) =  \frac {b_0} {2 B_0}  x  \, (1 -y^2)(1 -
 (z/z_m)^2) \exp(-3 y^2)   \hat {{\bf y}}.   
\label{xi}
\end{equation} 
Although this  displacement is even in $z$ this symmetry is
not transferred to components of the current density or the Lorentz
force.  
\begin{figure}
\centering
\includegraphics[width=2.0in]{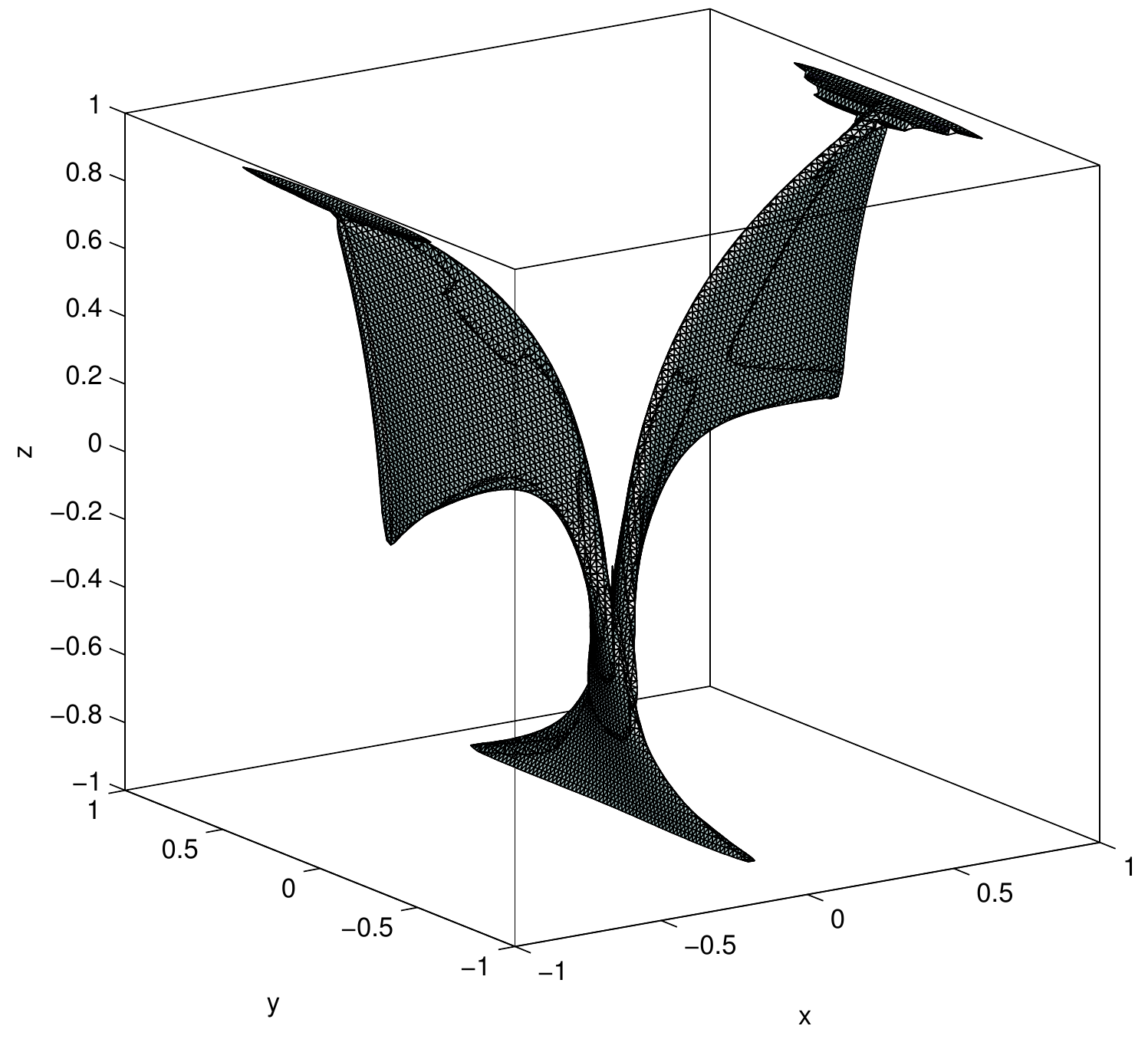}
\includegraphics[width=2.0in]{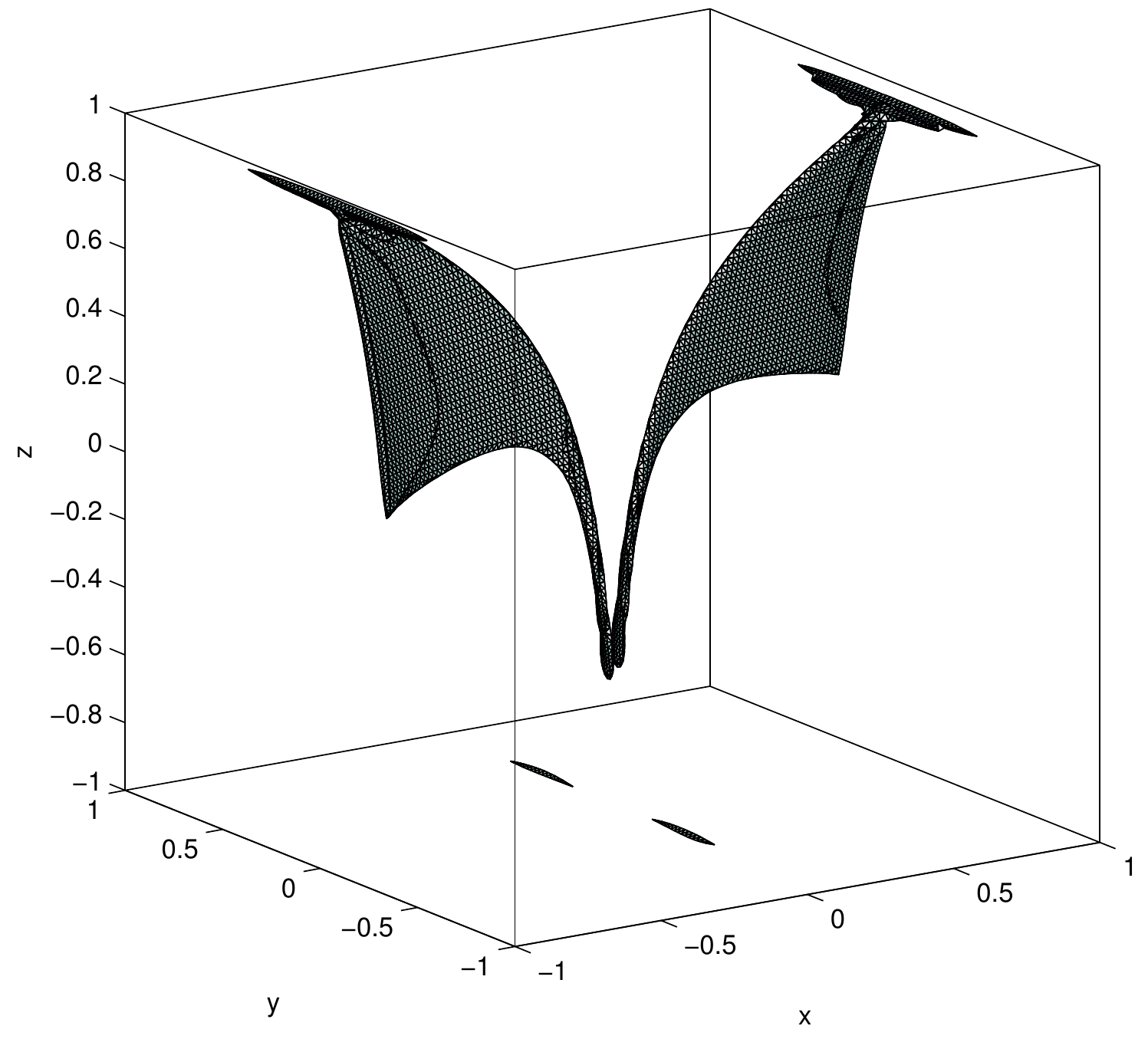} 
\includegraphics[width=2.0in]{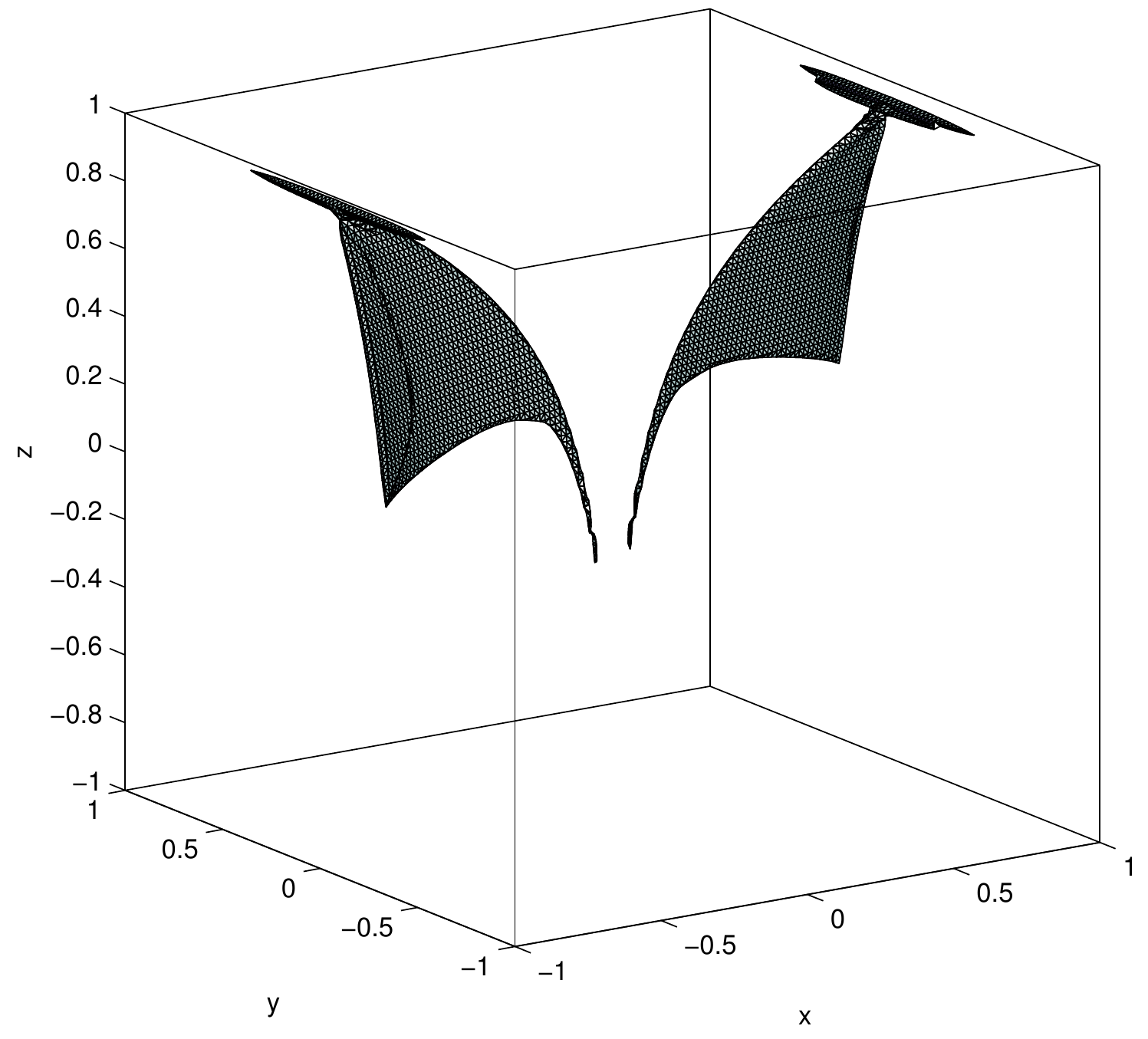} 
\caption {Surface current densities  for  half-length $z_m = 1$.   Shown left to
right are surfaces at  levels 1.3, 1.5,  and 1.6.   It is clear that
the highest concentrations of the current are limited to to the  
upper reaches of the QSL    Current accumulation along $z-$axis is
relatively minor.}  
\end{figure} 


\subsection{Relaxed current distributions}

The relaxation is performed using  the parameters $ B_0 = 1$ and $ b_0
= 0.6 $ for an axial field of strength $ B_z \, = \, 0.4$.    These
parameters,  though  comparable to those of   \cite{galsgaard2000},         
are chosen mainly to  highlight the role played by line-tying  
the axial field in determining  the distribution of current in the
relaxed configuration.    Fields are taken  to be 
``computationally relaxed'' when  the initial forces (of order unity)
have declined by over three orders of magnitude.   

Figure 1  illustrates the current structures obtained by
taking $z_m = 1 $ using  $N = 81$ support points in
each direction.     Shown are isosurfaces of the current modulus at levels $1.3, 1.5, 1.6 $
respectively.    It is clear that the higher current densities are
associated mainly with stresses in the vicinity of the upper line-tied
boundary $ z = 1 $.    In this example the axial field effectively 
prevents  the current localizing strongly towards the $z$-axis.
Notably these figures closely resemble  Figure 3 of \cite{galsgaard2000} who
uses the value $B_z=0.3 $ over a fixed computational  mesh 
(typically  $-0.5\leq x,y,z\leq 0.5$)  without considering in detail the
influence of the size of the domain.     

Figure 2 illustrates the results of a computation in which  $z_m = 2$.
The resolution of the previous  figure is maintained---mesh points in the
$z$-direction are increased in proportion to the extension in
height---and  all
other parameters are left unchanged.   The expectation is that the
greater tube height should weaken the influence of axial line-tying.   
Current surfaces taken at  $1.6, 1.8, 2.0 $ confirm 
that,  in contrast to  Figure 1,  current   localization is now
strongest  along the tube axis.    In this case the current
distribution delineates  mainly the lower reaches of the QSL.    

To interpret these results recall that,  in the case  of  infinitely long tubes 
($ \partial_z =  0$),   current  accumulation at  the  planar
null controls the reconnection rate in perturbed   
$X$-points.  This rate $ \eta  J $ is known to be ``fast''---invariant with
resistivity $ \eta $---only if $ {B_z}^2 /2 < \eta $
\citep{mcclymont1996}.   The presence of a finite computational domain, 
however,  changes matters radically since,  by 
line-tying the axial field,   QSL's are introduced.    It is these surfaces of steep
gradient in the field line mapping that now provide
sites for current accumulation.   But even in the
present rather simple geometry, the location of the peak current
\emph{within} the QSL is affected by a combination of
factors such as the length of the flux tube 
and the form of perturbation applied.   Note also,  that if the resistivity  were turned on, 
reconnection would involve field lines  
whose  end  points are anchored  across  different  $z-$planes,
as in  the field line `flipping' of \cite{priest1992}.    
\begin{figure} 
\centering
\includegraphics[width=2.0in]{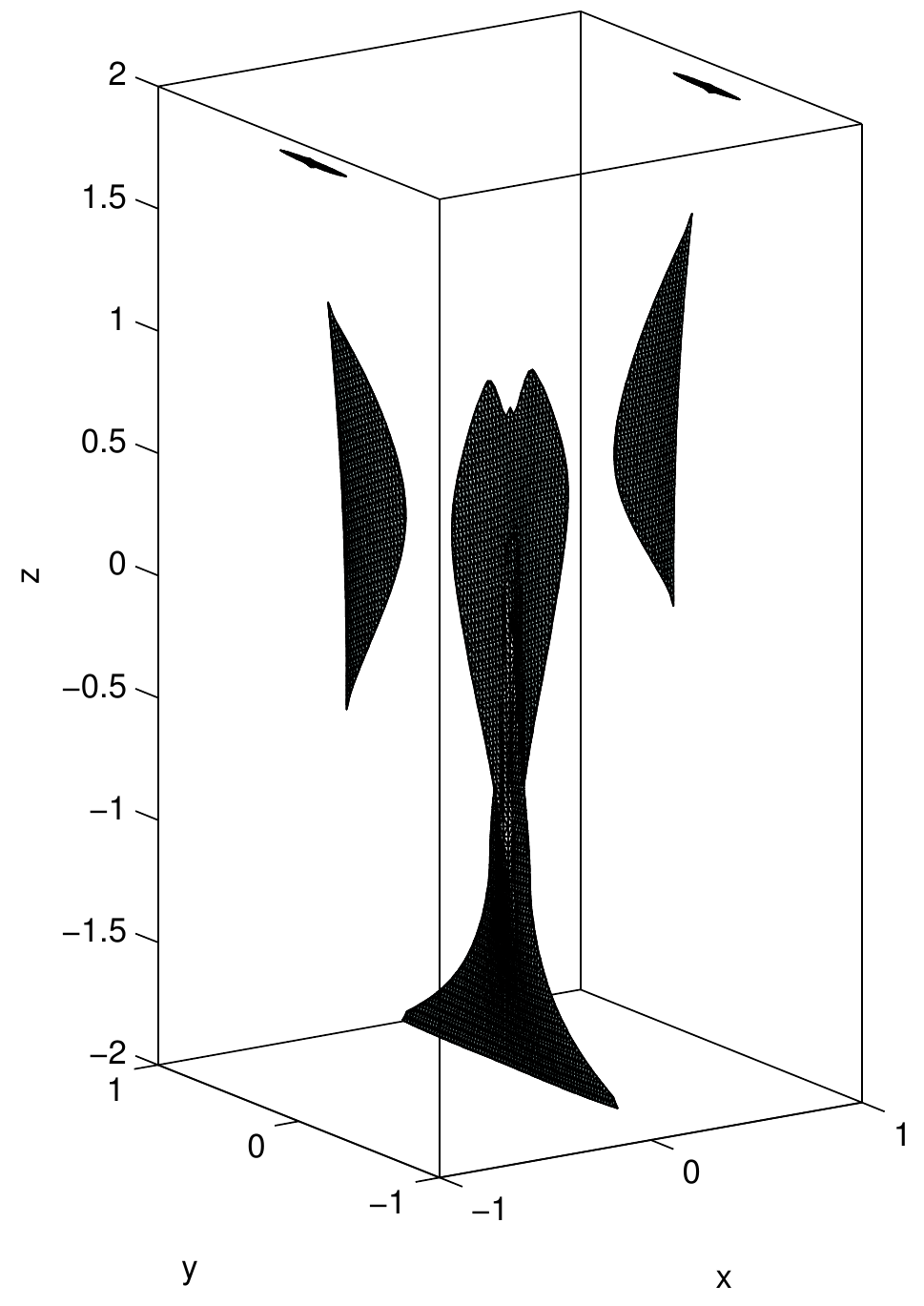}
\includegraphics[width=2.0in]{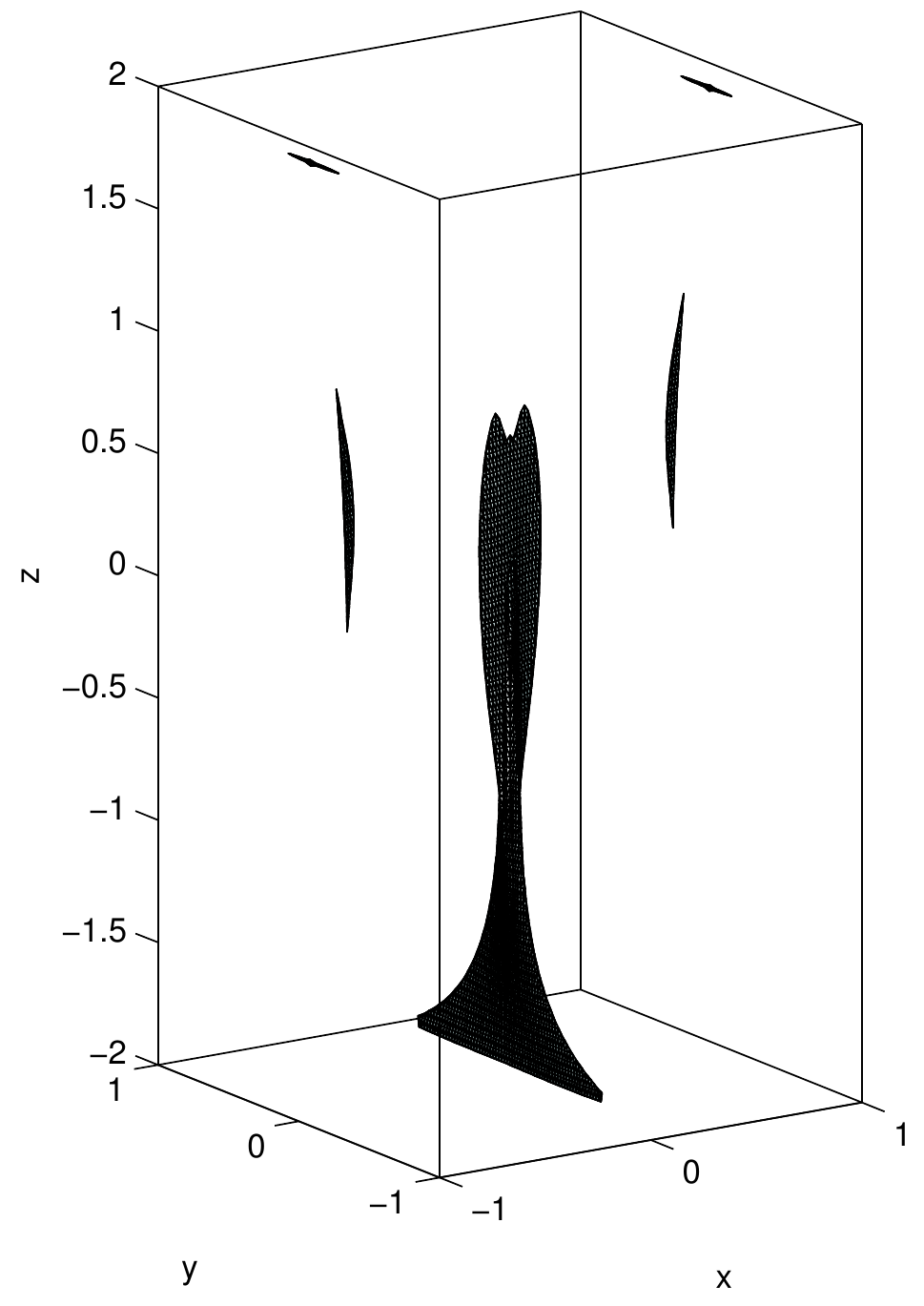}
\includegraphics[width=2.0in]{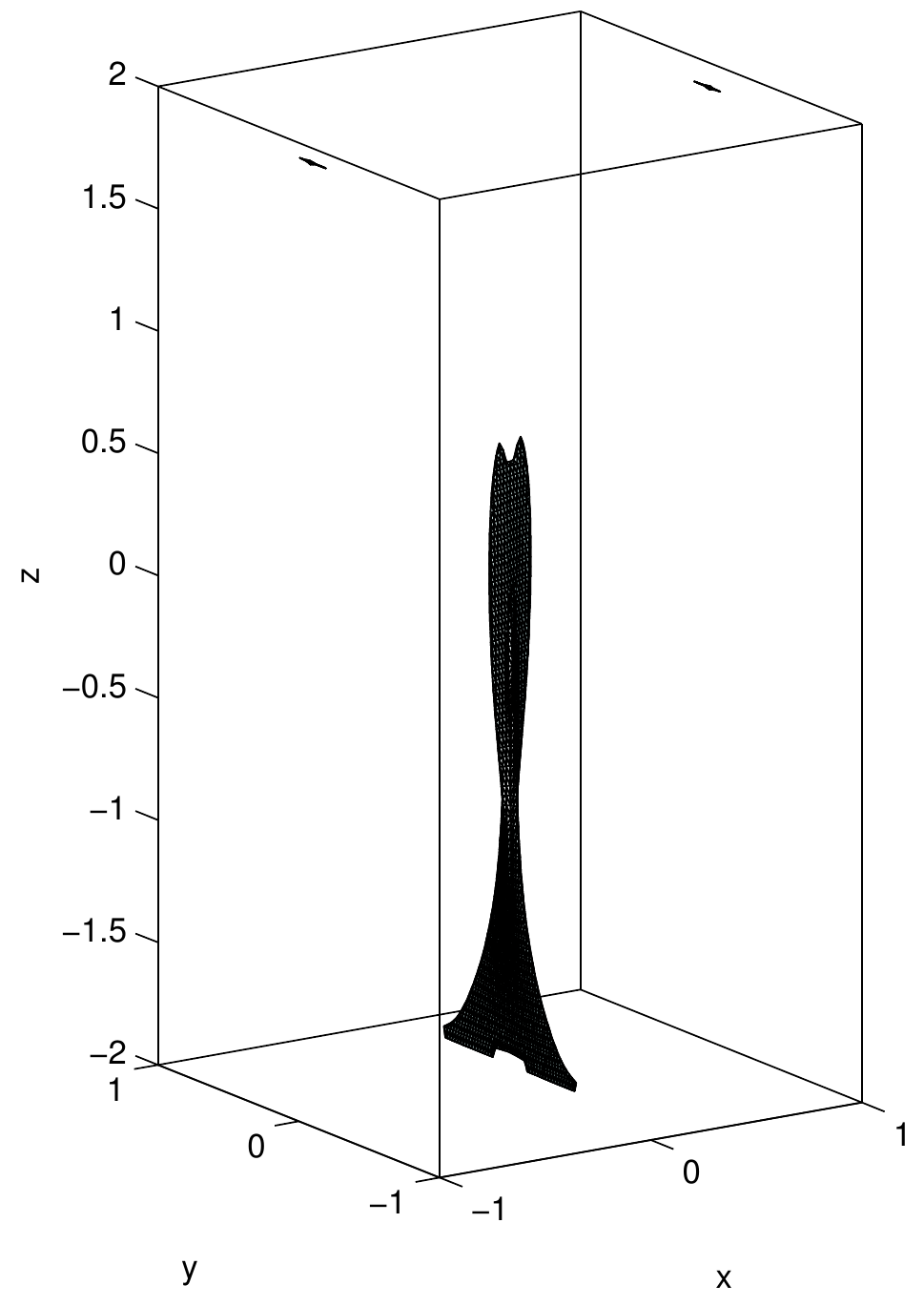}
\caption {Surface current densities  for  half-length $z_m = 2$.   Shown 
 left to right are surfaces at  levels 1.6, 1.8,  and 2.0.   In
 contrast to Figure 1,   it is now the lower levels of the QSL that
 are delineated by the current distribution.  }  
\end{figure}

\subsection{Current magnitude versus z-length}

We now examine,   more systematically,  how  the maximum current
densities on the mesh vary with changes in the height parameter $ z_m $.   
Figure 3 shows how  variations in $ z_m $ define three distinct
regimes.     For  $ z_m > 1.5  $  there is a regime in which the current density,   
localized along the vertical axis of the QSL (as in Figure 2), shows a
slow increase with increasing $z_m$.    
For  $ z_m < 1.2 $  the current  is concentrated towards   
the upper boundary footpoints  (as in Figure 1) and decreases with increasing
length.  The implication is that,  at sufficiently small lengths,  the
$B_z$-field is strong enough,  and line-tying effective enough,  
to prevent the current  localizing  around the $z$-axis.    
In the intermediate regime  $ 1.2 < z_m < 1.5 $ there is a transition that reflects the competition between the
interior and footpoint currents.  
\begin{figure}
\centering
\includegraphics[width=3.0in]{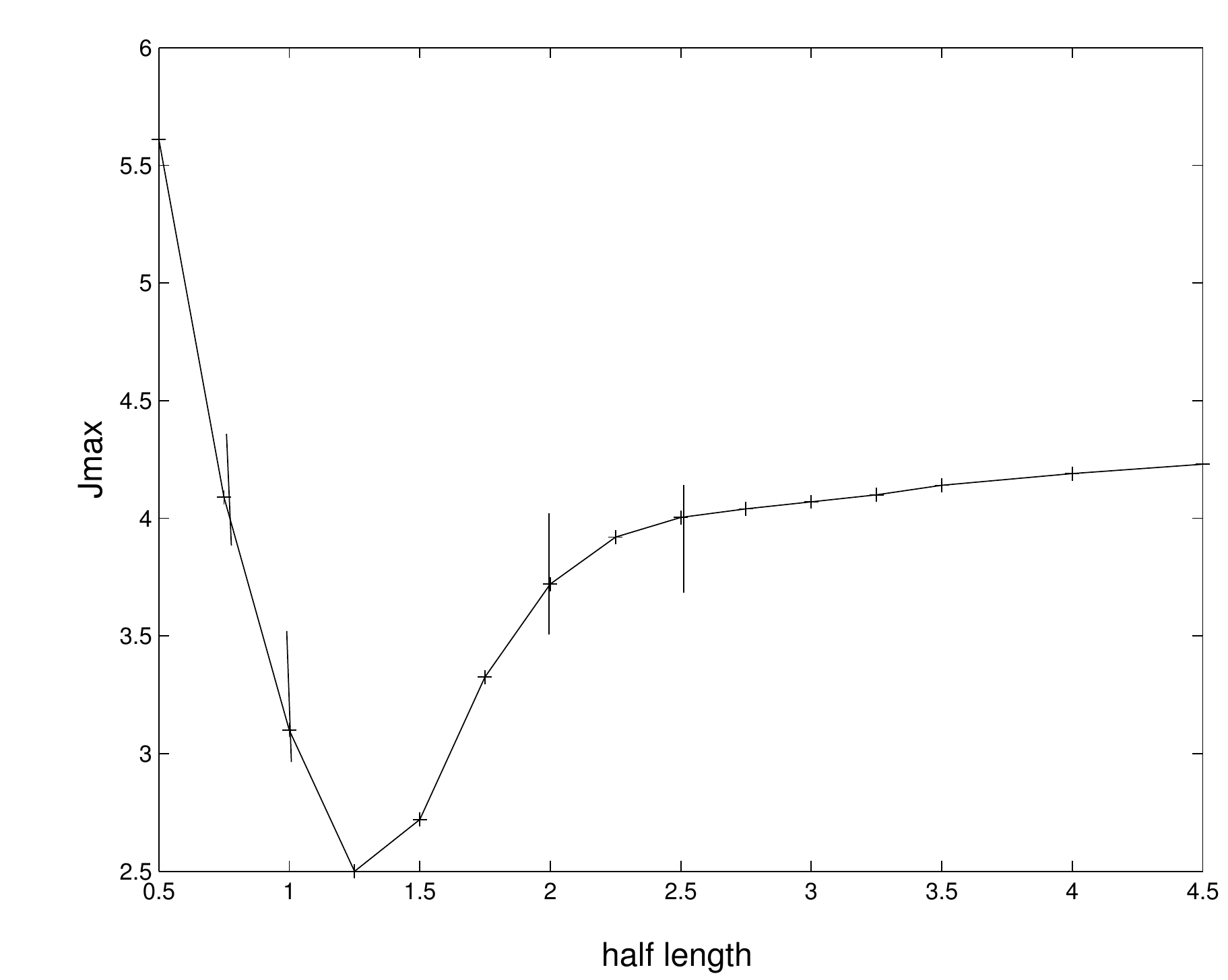}
\caption {Surface current density versus half-length $z_m$.   For
 sufficiently long tubes ($z_m > 1.5 $) the maximum current is located
on the axis of the QSL  and the peak current density slowly
increases  with tube length.    For $z_m < 1.5 $  the footpoint
currents associated with line-tying at the upper boundary become
significant.    These dominate  for sufficiently short tubes
and effectively prevent current localization on the tube axis.  } 
\end{figure}

It should be stressed that  the plots of Figure 3 are all computed
for a fixed numerical resolution (so height increases  are always achieved
by increasing  the mesh points in the $z$-direction). 
Independent of all other considerations,  however, 
the absence of resistivity means that,  in response to a topological
disturbance such as (\ref{xi}),   the current densities in the
relaxed configuration should  diverge as the resolution increases 
\citep{craiglitvinenko2005,pontincraig2005}.
If the current structure in the relaxed field is truly singular then, to the extent that $ B_z $ acts like an isotropic gas
pressure,  the current density is expected to approximate  a  power law  $ J \sim
N^\alpha $ whose exponent weakens with  the increasing strength of the
axial field  
\citep{craiglitvinenko2005}.
Line-tying $ B_z $ on the upper
and lower boundaries,  however,   introduces a magnetic tension force that  is
likely to slow this divergence.     {Of course,    the divergent behaviour could be quashed 
completely if,  at sufficiently high resolution,  the underlying current structure were
to have a finite thickness.    This occurs,  for example,   in a one-dimensional current sheet
halted by compressional effects due to gas pressure or finite $B_z$
\citep[][Appendix A]{craiglitvinenko2005}.}   

As already anticipated,   current tends to accumulate at the location
of  the QSL  in the field as identified by
\cite{priest1995}.    It is straightforward to show that for the
equilibrium field ($b_0=0$) the covariant 
squashing factor (between the planes $z=\pm z_m$) as defined by \cite{titov2007} takes the value 
\begin{equation}\label{qeq}
Q_\perp(x=0,y=0)=2\cosh \left(\frac{4z_m}{B_z}\right)
\end{equation}
on the $z$-axis. Clearly the squashing factor gets larger as $B_z$ is
decreased or $z_m$ is increased.  Therefore, in the regime $z_m>1.5$
when the current accumulates in the centre of the QSL, our 
results are consistent with the notion that the peak current in the
QSL increases as the squashing factor increases. While this behaviour
is expected on intuitive grounds,  it is  far from clear  that 
a direct link can be established between the
squashing factor in the initial field and the current intensity that
develops in response to the perturbation -- see \cite{demoulin2006} for a discussion.
   
\subsection{Divergence of the current magnitude with resolution}

Marked on Figure 3  are four vertical bars.   These indicate 
tube heights $z_m$ at which  we have explored the divergence of the current
density with resolution (as measured by $N$,  the number of grid points
across the tube axis).  Figure 4 confirms divergences that
approximate power law behaviour   for each of the
the four heights plotted.   Slight roll-offs are present for the
elongated tubes (say $ z_m \ge 2$),   but the shorter tubes---which
resist current localization   at  the centre of the tube---posses  stronger,   well defined
divergences ($ J \sim N^{0.83} $ for $ z_m = 0.75$).   
This suggests that,  independent of the tube length,  by line-tying
the $ B_z$ field we encourage current accumulation about the 
axial footpoints at the expense of current localization about the tube axis.
 
Finally in Figure 4 we present a  plot {($z_m = 1.0^*$)} that shows
 the convergence  obtained  when the tube is subject to a
 non-topological disturbance.   In this case the original perturbation
 (\ref{xi}) has been  modified  to provide a ``control''  disturbance that guarantees vanishing displacements at the origin
 and at the  tube footpoints.     In this case reconnection is not
 required  to regain the initial equilibrium and we therefore expect no localised current enhancements to develop.  
Consistent with this,  we see that,  in marked contrast to the
 divergence plots,   the convergence  of the current density (multiplied by
 fifty for graphical purposes) is accelerated  by increases in resolution.      
\begin{figure}
\centering
\includegraphics[width=3.0in]{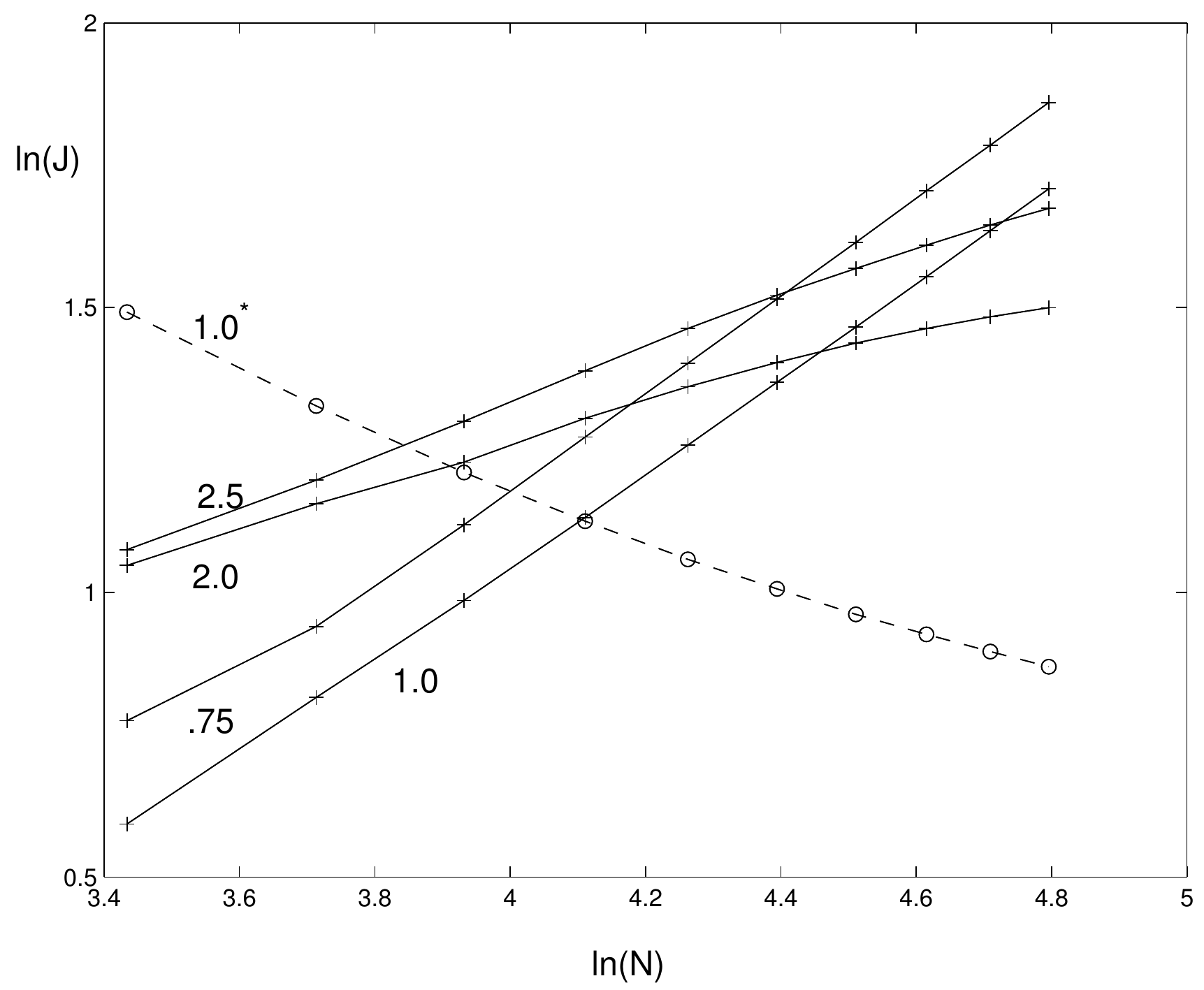}
\caption {{Divergence of the current density $J $ for various half-lengths
  $z_m$ as a function of the grid resolution $ N $.     Stronger divergences are obtained for (shorter) tubes which
  comprise localized footpoint currents.   Shown for comparison is a
  convergence plot  {(dashed line labelled $ z_m = 1.0^{*} $) } obtained using a non-topological disturbance for a tube
  of unit  half-length (with $J$ multiplied by
 fifty for graphical purposes).}} 
 \end{figure}

\subsection{The form of the perturbation}

It has already been noted that the current structures obtained by
magnetic relaxation are likely to be sensitive to the form of the applied 
perturbation.     Here  we briefly consider two variations of the disturbance
given by equation (\ref{xi}).      

\begin{figure}[ht]
\centering
\includegraphics[width=3.4in]{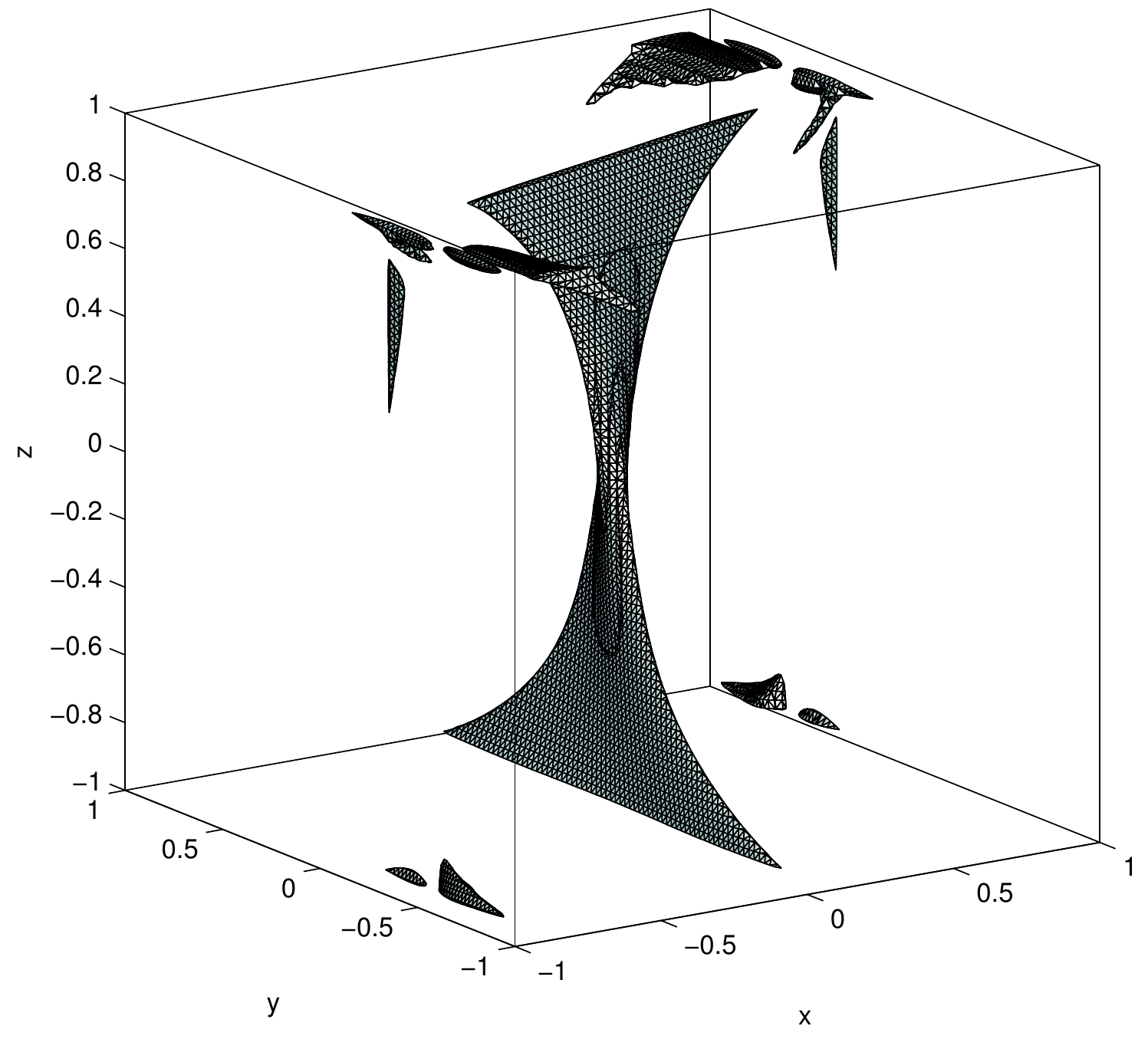}
\caption {Isosurface of the current modulus at $36 \%$  of the 
 maximum for a simulation in which the applied perturbation is
 non-zero on the $z$-boundaries -- see equation (\ref{xi_nozdep}). 
 Despite the well represented QSL layer in the present  figure ($ z_m = 1.0  $),
 we find  that the QSL axis currents can dominate
 footpoint currents only when the tube is sufficiently long ($ z_m > 1.5 $).}
\label{jsurf_nozdep}
\end{figure}

Figure \ref{jsurf_nozdep} shows the current structure  obtained  by turning off the $z-$dependence  of  the
perturbation, specifically by taking
\begin{equation} \label{xi_nozdep}
 {\bf \xi}(x,y,z) =  \frac {b_0} {2 B_0}  x  \, (1 -y^2)\exp(-3 y^2)   \hat {{\bf y}}.   
\end{equation} 
The entire QSL is well  represented---compare Figure 1
for $ z_m = 1$---but localized currents attached
to the footpoints  are also clearly present.   Although it is tempting to interpret these
footpoint  currents as numerical artefacts,  it  should be remembered that real magnetic
stresses are present on the line-tied boundaries  and that,
according to the present computations,    these stresses
become increasingly  dominant (compared with currents on the QSL axis)
for the shorter tubes.   Computations based on  those in  \S2.3 again confirm that  only 
when the effects of  line-tying are weakened,  specifically by taking $
z_m > 1.5 $,   is a dominant current distribution localized to the
axis of the QSL realized.

\begin{figure}[ht]
\centering
\includegraphics[width=3in]{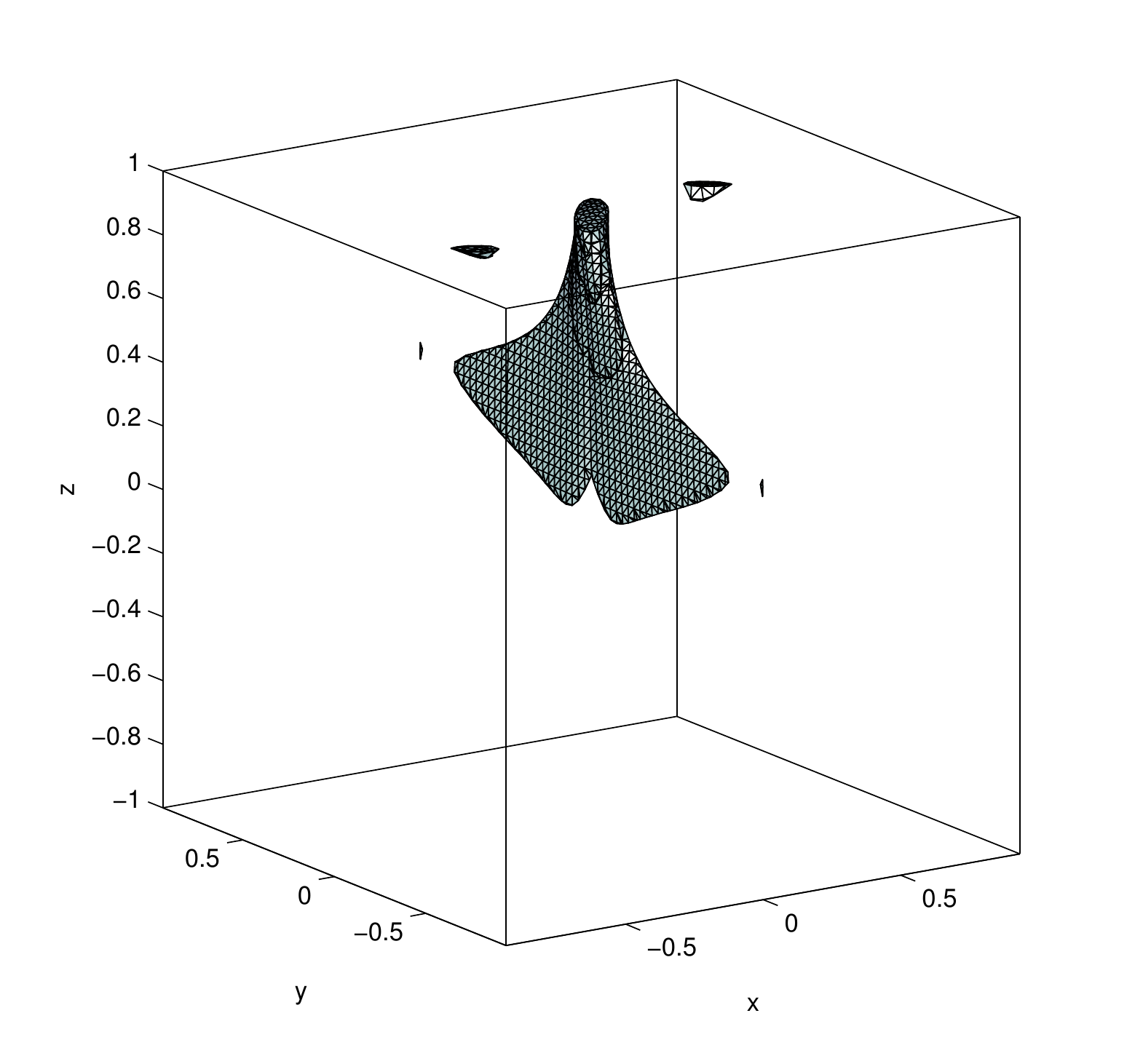}
\includegraphics[width=3in]{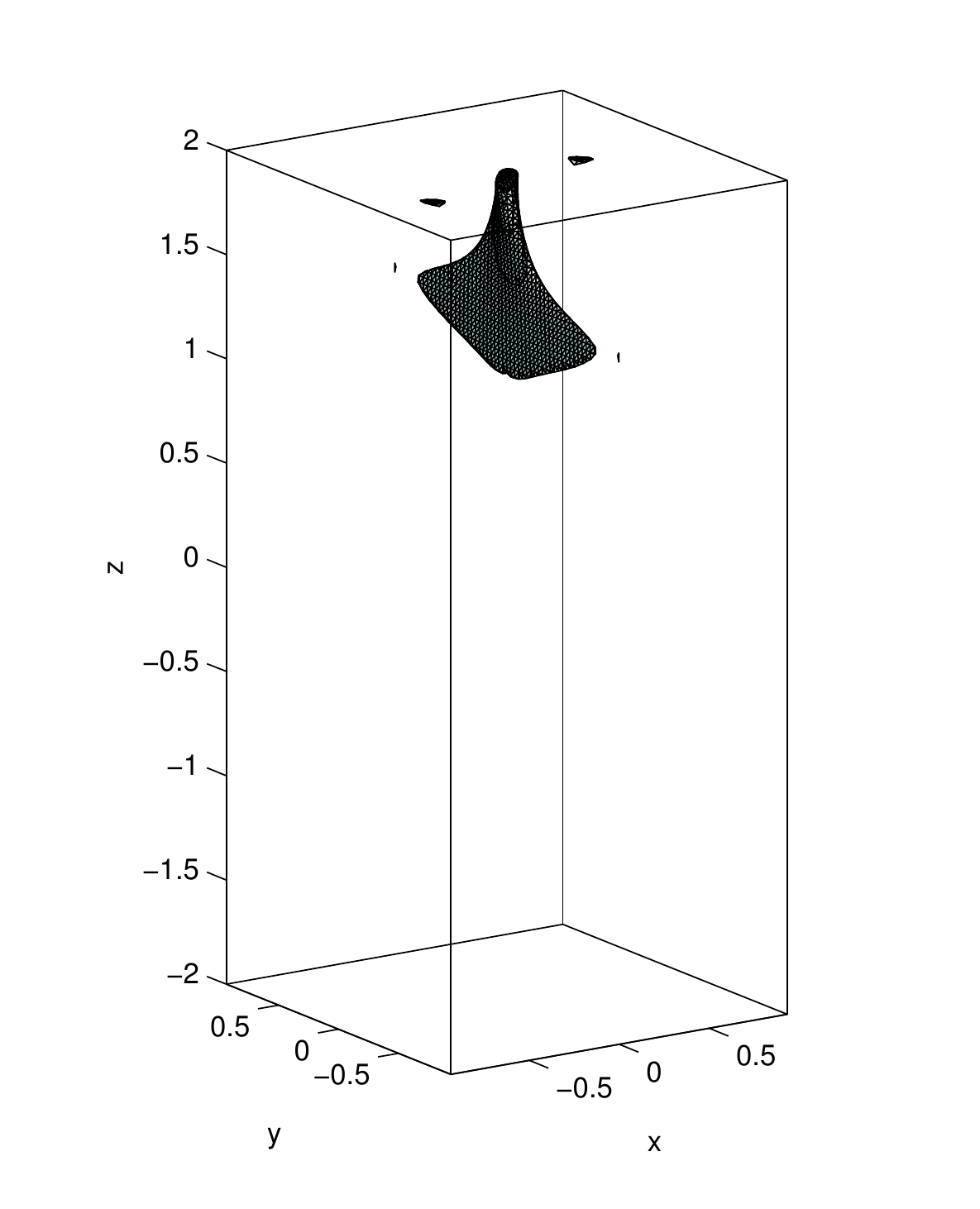}
\caption {Isosurface of the current modulus at 70\% of maximum for two simulations in which the applied perturbation is zero on the lateral boundaries but non-zero on the $z$-boundaries -- see Equation (\ref{xi_zbouonly}) -- for $z_m=1$ (left) and $z_m=2$ (right). }
\label{jsurf_zbou}
\end{figure}

Finally,   in Figure \ref{jsurf_zbou},   we show the result of
adjusting the disturbance such that it is zero on all lateral
boundaries. This would be equivalent in an MHD simulation to eliminating any 
driving from the lateral boundaries, but driving only from the $z$-boundaries. Specifically, we take
\begin{equation} \label{xi_zbouonly}
 {\bf \xi}(x,y,z) =  \frac {b_0} {2 B_0}  x  \, (1 -y^2)(1-x^2)\exp(-3x^2-3 y^2)   \hat {{\bf y}}.   
\end{equation} 
In this case the 
perturbation is non-zero only on the top and bottom surfaces (which
in previous studies of the QSL geometry were often identified as
fragments of the solar photosphere \citep{inverarity1997,galsgaard2000}).  
The interesting aspect of the resultant  current structure  is that,
in being concentrated to the upper regions of the tube,  
the distribution is rather insensitive to increases in the tube length (especially
compared with Figure 3).  That is, there appears to be a limit to
the distance along the tube that 
the current can penetrate.  The peak current diverges with resolution
as before,   but the  scalings for tubes longer than $ z_m \ge 1 $
show only minimal variations.  This is consistent 
with the results of \cite{galsgaard2000}, who found little variation
in the current structure obtained through a dynamical driving when the
domain was doubled in length. On the other hand, for 
tubes with $z_m \ll 1$, the current appears to converge to a finite
value, indicating that the underlying 
current layer is probably of finite thickness.  This behaviour
clearly differs  from our previous examples and may derive from  the fact that  the perturbation  no longer involves
displacements of the lateral footpoint boundaries.   We return to a
discussion  of the possibility of non-singular QSL currents in \S4 below.

\section {3D null points} \label{3dsec}

\subsection{The simulation setup}

We now explore the current structures obtained by perturbing,
non-linearly,   a fully 3D equilibrium field.     Of particular
interest is the extent to which the results of the 
line-tied $X$-point model of
Section 2 can be applied to more general  3D magnetic nulls.    
An exact equivalence cannot be expected since,  in the absence of
 a true magnetic null,  all points in the QSL field considered above are
causally related.    We present only a preliminary study here,  but  point out that the
competition  between localized interior currents and 
 {the footpoint currents},   so apparent in the  previous analysis,
is also a salient feature in the model outlined below.    
For the cases considered   however,   the current at the null is
always  ultimately dominant.      

Our initial condition is a potential magnetic field due to four flux
patches on the lower boundary of the domain, $z=-z_m$. The field is
generated by four point charges that lie outside the domain at $z<-z_m$. The resultant magnetic field is given by 
\begin{equation}\label{b_pot}
\BB_p=\sum_{i=1}^4 \epsilon_i \frac{\xx-\xx_i}{|\xx-\xx_i|^3}
\end{equation}
where $\xx_i$ are the locations and $\epsilon_i$ are the strengths of
the point charges. Here we take
$\{\epsilon_1,\epsilon_2,\epsilon_3,\epsilon_4\}=\{-0.5,-0.5,1.05,-0.015\}$ 
and $\xx_1=(0,0.6,-1.0),~\xx_2=(0,-0.6,-1.0),~\xx_3=(0,0,-1.3),~\xx_4=(0,0,-0.7)$.
In what follows we therefore restrict the domain height parameter to $z_m<0.7$.
\begin{figure}[t]
\centering
\includegraphics[width=4in]{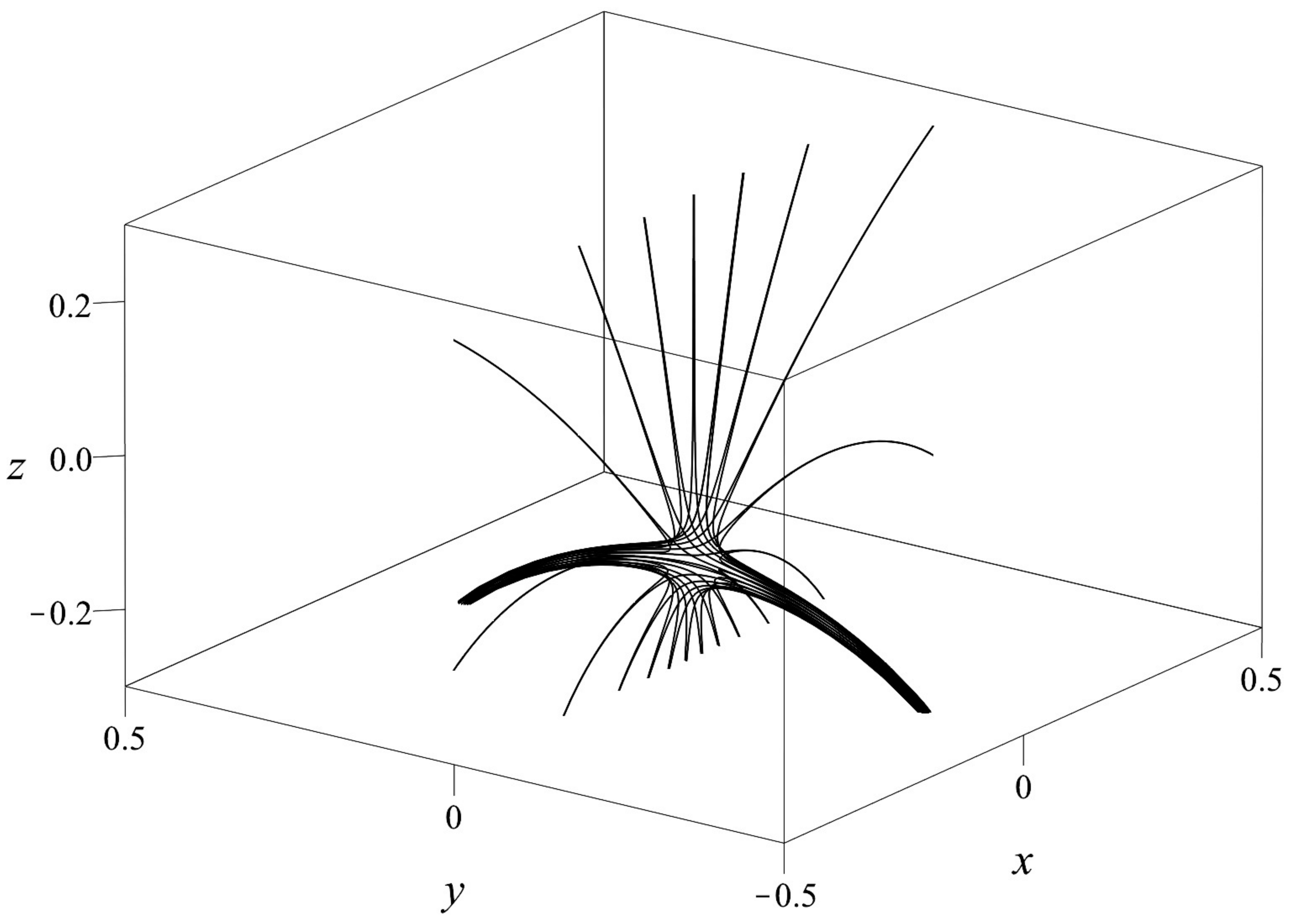}
\caption{Spine-fan structure of the equilibrium magnetic field containing a 3D null point, with $z_m=0.3$.}
\label{B3dfig}
\end{figure}
The field structure is shown in Figure \ref{B3dfig}, and possesses a
well defined spine and fan structure that emanates from
the isolated  null at $ x = y = 0,   z = -0.18 $.  

The geometry of the magnetic field is significantly more
complicated than the linear null considered in the previous study of
\cite{pontincraig2005}, since the spine 
of the null curves down to intersect with the lower boundary.  However,
the advantage of this  is 
that all field lines that pass close to the null are anchored at
$z=\pm z_m$ at least at their spine end (and in many cases at
both ends).   As such we can hold 
the domain dimensions in $x$ and
$y$ fixed since the line-tying there is relatively passive in the
formation of currents at the null, and make a more direct comparison
with the results of Section 2.  Regarding comparison with the study of
\cite{pontincraig2005},  it 
is worth noting that the field in the
fan plane of the null is relatively isotropic: the ratio of the fan
eigenvalues is 0.84.  

We perturb our 3D equilibrium  by adding the displacement   
\begin{equation} 
 {\bf \xi}(x,y,z) =  b_0 x  (1 -y^2)(1 - (z/z_m)^2) \exp(-3 y^2)  \, 
 \exp(-10 x^2)   \, {\hat {\bf y}}.
\label{dist3d}
\end{equation}  
 This perturbation has the effect of deforming the field in the
 vicinity of the null such that the spine 
and fan are no longer perpendicular. This leads to a Lorentz force
that acts to increase the null collapse, as discussed in \cite{pontincraig2005}.
The resulting non-equilibrium is then allowed to dynamically relax in 
the interior   ($  -1 <  x,y < 1,  |z| < z_m  $) 
while remaining line-tied on all boundaries.

\subsection{Localization of the current density}

It is interesting that the 3D null field,  when
perturbed according to (\ref{dist3d}),   has similar current
localization properties to the planar null of Section 2.    Notably
the current accumulates either at the null or towards the 
lower boundary in the vicinity of the footpoints of the spine and fan field lines. 
Which of these competing distributions is dominant  can be
expected to depend both on the height of the domain and the resolution
of the computation, as well as 
the form of perturbation used.

\begin{figure}[ht]
\centering
(a)\includegraphics[width=3.6in]{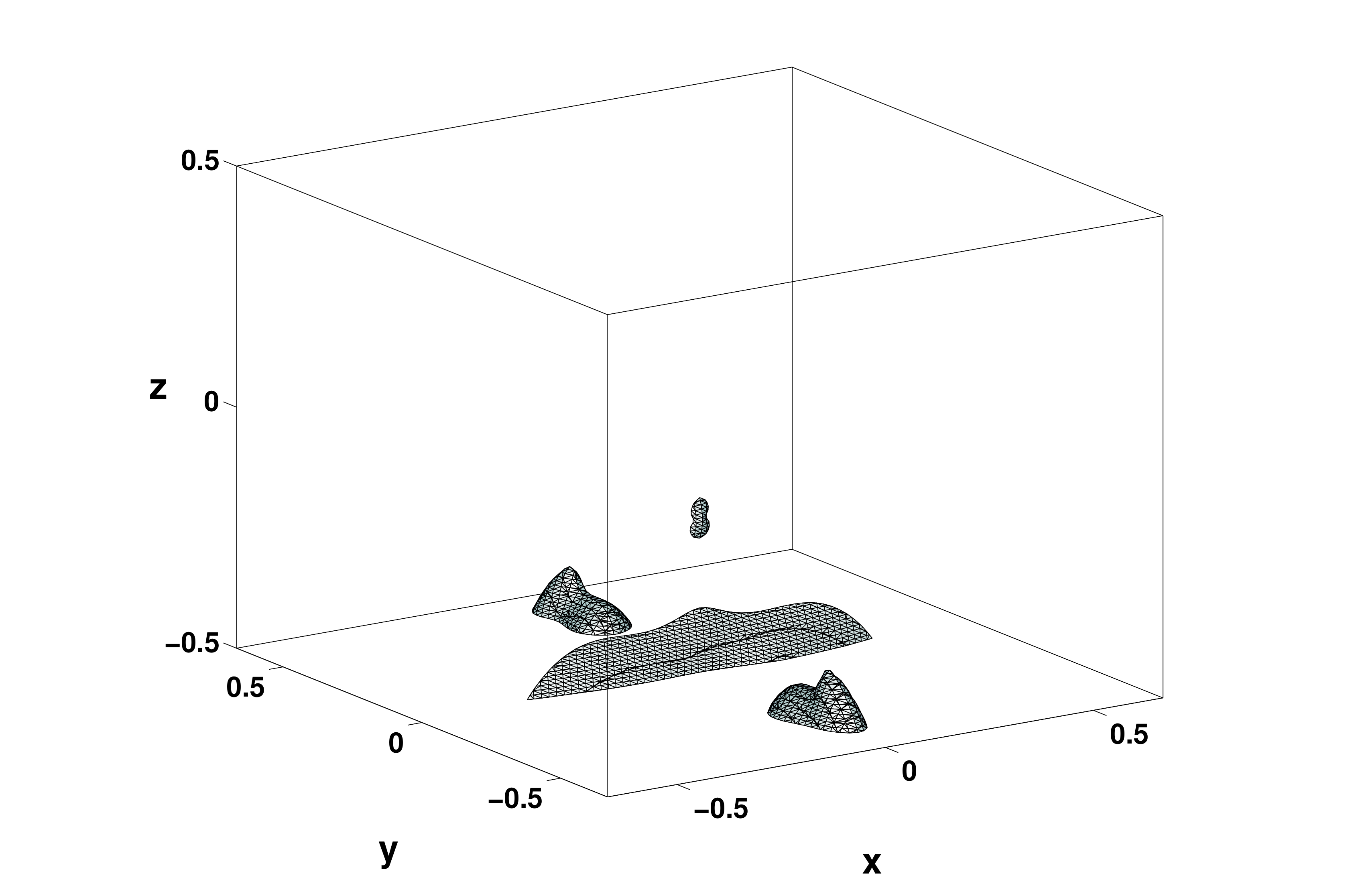}
(b)\includegraphics[width=3.6in]{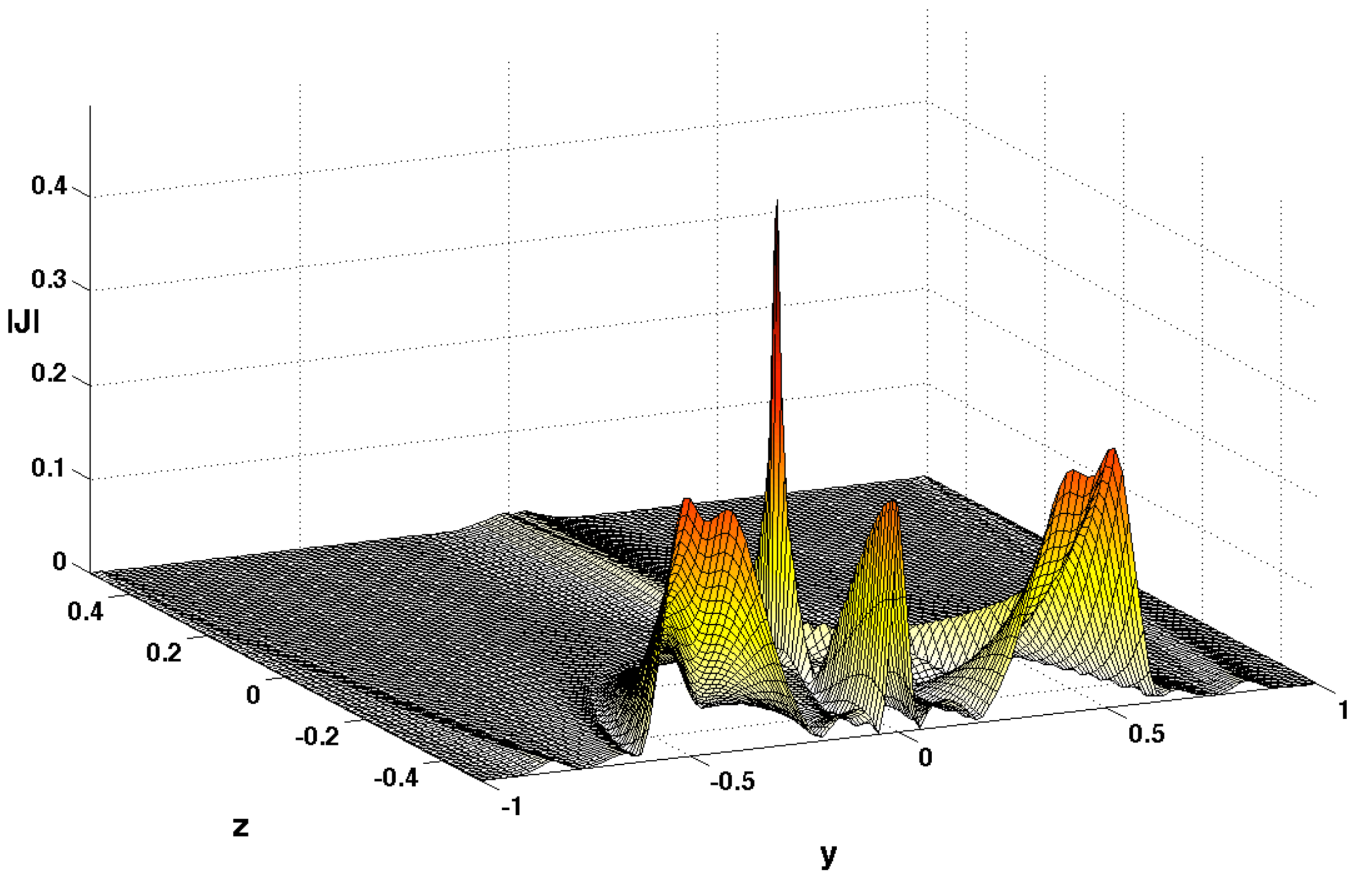}
\caption {Current density $|{\bf J}|$ in the relaxed state of the
field with the 3D null point. 
Mesh resolution is $121^3$ and $z_m=0.5$. (a) Isosurface of current at
25\% of the maximum in the domain. 
(b) Surface plot of the current density in the mesh plane initially coincident with the $x=0$ plane.} 
\label{J3dfig}
\end{figure}

\begin{figure}[t]
\centering
\includegraphics[width=4.0in]{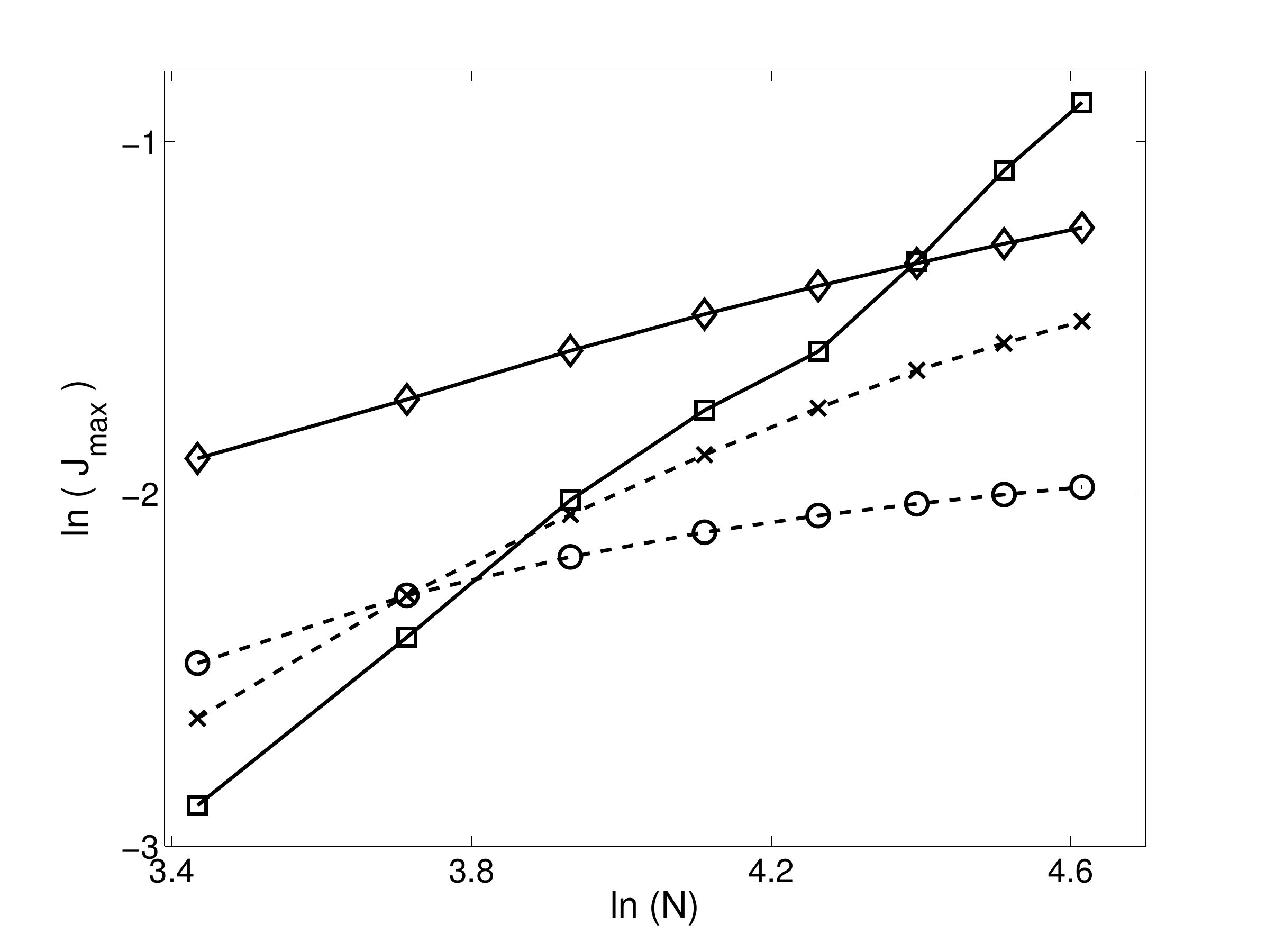}
\caption {Scaling of the peak current density {$J$} for $z_m=0.5$ (solid
lines) and $z_m=0.3$ (dashed lines). Plotted are 
the current at the null (squares for $z_m=0.5$, crosses for
$z_m=0.3$), and in the boundary 
current concentrations (diamonds for $z_m=0.5$, circles for $z_m=0.3$).} 
\label{3dscalings}
\end{figure}
Figure \ref{J3dfig} shows a typical current localization for the case $ z_m =
0.5 $.     Note that in contrast to the discussion of Section 2,   we
now find it instructive to  measure {\it both} the peak current at the null point and the peak
current in the footpoint regions.    Figure \ref{3dscalings}  compares the null and footpoint current distributions as a
function of resolution in the case $ z_m = 0.5$.    In common with the planar nulls of Section
2  (see e.g.~Figure 4),   the current at the null eventually dominates,  
approximating a strong power law divergence $ J \sim N^{1.64} $.    In
this case  the current at the spine and fan footpoints is strongest at low
resolution $ J \sim N^{0.58} $,    but  is readily overtaken as the
resolution is increased.   

We have also repeated the above computation for the
case $ z_m = 0.3 $.  If we focus on providing a significant  ``reconnection
current''   at  the null,  then we would expect that current densities should weaken as the tube shortens (in
the manner of regime 1,  $ z > 1.5 $,  of Figure 3).    However, what needs to
be borne in mind is that lengthening the domain leads to larger field
intensities, and therefore potentially larger currents, near the base of the
domain.  In this case it therefore makes sense to consider not the absolute value of
the current density but its scaling with resolution.   
For the simulations with $z_m=0.3$,   line tying should now more
strongly  resist the collapse to singularity at the null, and so we expect a weaker scaling of the peak current with resolution $N$.
As compared to Figure \ref{3dscalings}(a),  the blow up of the current density is indeed noticeably
weaker, $ J \sim N^{0.94} $,  and the roll-over of the boundary current scaling even more pronounced.    
Importantly, it appears  that for the null point field the reconnection
 current always eventually dominates,  for sufficiently high
 resolution,  in contrast (say) to the $ z_m =.75$  and $ z_m = 1$ results for the linear $X$-point.  

\section {Discussion and conclusions}

We began by considering the action of a topological disturbance on a 
2D $X$-point threaded by a uniform axial field $B_z$.     The initial field is
line-tied on all boundary surfaces, and contains  
no magnetic null but rather a quasi-separatrix layer (QSL) aligned to the tube axis (here the $z$-axis).
The absence of resistive effects means that  very large currents
can be expected to develop,   in response to displacements of the
lateral footpoints,   over regions  delineated by the QSL.  
%
%

The literature appears divided  over whether   
one should should expect current densities that are formally
singular within the QSL.    Certainly, in the limit of an infinite domain 
length (corresponding to $Q_\perp\to\infty$, see Equation (\ref{qeq})), 
we expect a singular layer to develop 
\citep{mcclymont1996,craiglitvinenko2005}.   Notably,   working in a
finite-length domain,  a  logarithmic current singularity
was obtained by \cite{inverarity1997} for a particular form of boundary driving.    
Other studies,   in  contrast,    propose that,   in response to a dynamic driving from
the boundaries,   the current  in the QSL should form an intense but finite layer whose thickness is
in some way related  to the thickness of the QSL
\citep[e.g.][]{titov2003,galsgaard2003,aulanier2005}.   This proposal  
was recently examined by 
\cite{effenberger2011}  who used an AMR  code to  follow the current collapse in a hyperbolic flux
tube (defined by the intersection of two QSL's).  Even
at very high levels of grid refinement,  no saturation of the current
layer was observed.   Indeed,  while intuitively one expects the intensity of the current in the
layer that forms in the QSL to be proportional to the squashing factor $Q$, a concrete
theoretical demonstration remains elusive, 
as discussed by \cite{demoulin2006}. What is more, the particular
location {\it within} the QSL that most 
favourably attracts the current has not received much attention
(though see Effenberger et al 2011)  except
in a handful of highly symmetric geometries. 

In the ideal relaxation simulations described in \S2, we find that for
axial fields  $B_z$  of comparable strength to
the planar field,   magneto-frictional  relaxation  provides
two distinct outcomes for the relaxed current distribution,
depending on the length $ 2 z_m$  of the tube and the form of the
perturbation.  If the tube is short enough ($ z_m < 1 $,  for the examples studied),   current localization on the tube axis 
is inhibited,   and currents  become attached mainly to  the line-tied
footpoints  on the upper and lower surfaces.   This behaviour is not
likely to be favourable to the onset of strong magnetic reconnection.  

Current accumulation on the axis of the QSL can occur
however,  despite the influence of line-tying,  provided the  tube is
sufficiently long.  For $ z_m > 2 $  currents  on the tube axis can completely detach from the upper
and lower boundary  surfaces.      Longer tubes now   
lead to larger axial  currents and potentially  faster reconnection.
Although  the relaxed current distribution  becomes near-singular,
the smallness of the plasma resistivity in coronal plasmas means that 
frictional collapse should provide  a useful guide to the strength and
morphology of 3D reconnective currents.    This view is supported by the transient resistive computations
of \cite{galsgaard2000}   which provide current structures consistent
with QSL's  that  closely resemble the present results.   

Whether the current sheet obtained after an ideal relaxation is
formally singular is difficult to determine from a numerical simulation. A current layer of finite width would 
be expected to show a ``roll-over" in the peak current divergence with
resolution---that is,  for sufficiently high resolution the peak current
should converge  yielding  a  force free solution comprising sharp but
non-singular current layers.  What can be said is that,  for the
limited resolution available to us,   and for all simulations in which the
lateral boundaries are perturbed,   we have observed no sign of  any
such roll-over.  Only in the case of short tubes (with low $Q$),
perturbed only on the $z-$boundaries,  do 
we observe current convergence indicating an underlying layer of finite width.  

In Section 3 we explored line-tying in the context of  a  perturbed 
3D magnetic null.    In this case the null point,   as opposed to the
QSL of Section 2,    is able to 
provide a unique focus for the magnetic stresses.    Relaxation computations
confirm that current accumulates  not only at the null
but also in the footpoint regions associated with  the line-tied
boundaries.   Null-point currents,  however, 
are found to be  the  most intense. Indeed, a collapse to a singular
current layer at the null is indicated in all simulations, despite 
the inhibiting influence  of line-tying which acts to weaken the
divergence of the peak current with resolution.       
Accordingly, when  the effects of line-tying  are weakened---essentially  by increasing  in the
height of the null point above the line-tied boundary---the  reconnective currents at the
null are significantly enhanced.   Given  the strong interest in
factors affecting the speed of the magnetic reconnection,  it will be of some    
interest to check the veracity of these findings in more general,
weakly resistive,  3D plasmas. 


\smallskip
    
\section*{Acknowledgements}

Comments by Yuri Litvinenko have been much appreciated. D.P. acknowledges the financial support of a Philip Leverhulme Prize.


\end{document}